\documentclass[a4paper,prb,showpacs,superscriptaddress,twocolumn]{revtex4}
\usepackage{amsfonts}
\usepackage{amssymb}
\usepackage{amsmath}
\usepackage{graphicx}

\usepackage{psfig}
\usepackage{bm}

\begin{document}

\title{Arbitrary rotation and entanglement of flux SQUID qubits}

\author{Zsolt Kis}

\affiliation{Research Institute for Solid State Physics and
Optics, P.O. Box 49, H-1525 Budapest, Hungary}

\author{Emmanuel Paspalakis}

\affiliation{Materials Science Department, School of Natural
Sciences, University of Patras, Patras 265 04, Greece}

\date{\today}

\begin{abstract}
  We  propose  a  new  approach   for  the  arbitrary  rotation  of  a
  three-level SQUID qubit and describe a new strategy for the creation
  of coherence  transfer and entangled states  between two three-level
  SQUID   qubits.   The  former   is   succeeded   by  exploring   the
  coupled-uncoupled  states of  the  system when  irradiated with  two
  microwave pulses, and  the latter is succeeded by  placing the SQUID
  qubits into  a microwave cavity  and used adiabatic  passage methods
  for their manipulation.
\end{abstract}


\maketitle

\section{Introduction}
It has been realized over the  last few years that solid state systems
that make use of the Josephson  effect could play an important role in
the  area  of  quantum  computation  \cite{Makhlin01a}.  A  series  of
successfully  performed interesting   experiments 
\cite{Nakamura99a,Mooij99a,Friedman00a,Han01a,Vion02a,Yu02a,
  Martinis02a, Wallraff03a, Pashkin03a, Chiorescu03a, Berkley03a} have
confirmed the  applicability of these  systems.  Among superconducting
systems performing  quantum computations,  emphasis has been  given to
the study of schemes based  on magnetic flux states in superconducting
quantum  interference  devices  (SQUID)  \cite{Makhlin01a,  Bocko97a,
  Chiarello00a,  Averin00a,  Everitt01a,  Zhou02a,  Amin02a,  Yang03a,
  Yang03b}.

Zhou {\it et al.}  \cite{Zhou02a} have recently proposed a three-level
$\Lambda$-type rf-SQUID qubit.  Here, the  states of the qubit are the
two lower flux states $|0\rangle$  and $|1\rangle$ of the SQUID system
that   reside  in   two   distinct  potential   valleys,  see   figure
{\ref{fig1}}.   The  manipulation  of  the  qubit  is  done  with  two
microwave  fields that  couple  the  lower states  to  an upper  state
$|e\rangle$  in a  $\Lambda$  configuration.  As  the coupling  matrix
elements corresponding  to the transitions  $|0\rangle \leftrightarrow
|e\rangle$ and  $|1\rangle \leftrightarrow |e\rangle$  are larger than
that  of  the $|0\rangle  \leftrightarrow  |1\rangle$ transition,  the
three-level SQUID qubit  has been shown to be  more favorable than the
conventional two-level SQUID qubit.  Amin {\it et al.}  \cite{Amin02a}
have  latter  shown  that the  approach  of  Zhou  {\it et  al.}   was
incomplete and  have proposed a  more general method for  producing an
arbitrary  qubit  rotation  using  the three-level  SQUID  qubit.   In
addition, more  recently Yang and  Han \cite{Yang03a} have  shown that
large detuning  of the  driving fields from  the upper state  could be
favorable for  implementing single  qubit rotation in  the three-level
SQUID qubit.  Finally, Yang  {\it et al.}  \cite{Yang03b,Yang03c} have
proposed two different strategies  using three-level SQUID qubits in a
microwave cavity  for entanglement, logical quantum  gates and quantum
information transfer.

In this article our goal is two-fold. First, we propose a new approach
for the  arbitrary rotation of  a three-level SQUID qubit.   The basic
idea  of the  rotation procedure  described  here is  related to  that
studied by one of the authors \cite{Kis02} for atoms driven by optical
fields.   However, there  the rotation  is  performed by  means of  an
adiabatic  process, that  requires  the involvement  of an  additional
long-lived auxiliary level  and an extra coupling field.  Here, we use
only two coupling  fields and present a rotation  method based on Rabi
oscillations.  As we  are in  the microwave  domain,  the sufficiently
precise control of  the pulse area is possible,  unlike in the optical
domain.  Second,  we apply  a  new  strategy  for the  realization  of
coherence transfer  and for the  creation of entangled  states between
two three-level SQUID qubits.  This is succeeded by placing the qubits
into a  microwave cavity  and by using  adiabatic passage  methods for
their manipulation. Both of our approaches are fundamentally different
from the ones that have been proposed so far in the literature as they
are based  on the exploration of  the coupled and  uncoupled states of
the system.  Moreover,  the latter method is an  adiabatic method, and
such  methods  are robust  with  respect  to  fluctuations of  several
experimental parameters  \cite{Vitanov01}. We will  discuss this issue
for our case below.

The article  is organized as follows.  In the next  section we briefly
summarize  the model  for the  three-level $\Lambda$-type  SQUID qubit
\cite{Zhou02a} and  present our method  for arbitrary qubit  rotation. 
The general method  is also specialized in two  simple cases. Then, in
section III we  study the interaction of two  three-level SQUID qubits
in  a microwave  cavity and  show  that using  adiabatic methods  both
robust  quantum information  transfer and  entanglement are  possible. 
Finally, we summarize our results in section IV.

\section{Rotation of a three-level SQUID qubit}

\begin{figure}
\begin{center}
\centerline{\hbox{ \psfig{figure=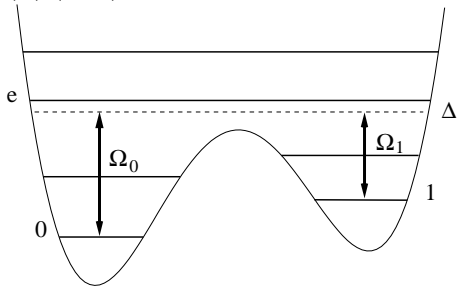,width=6cm}}}
\end{center}
\caption{The potential energy, the stationary states and the coupling 
  configuration  of  the  rf-SQUID  system.  The  states  $|0\rangle$,
  $|1\rangle$, and  $|e\rangle$ are coupled my means  of two microwave
  pulses   with  Rabi  frequencies   $\Omega_{0,1}$  in   a  $\Lambda$
  configuration. } \label{fig1}
\end{figure}

The model qubit consists of an rf-SQUID which interacts with  two
microwave fields. The rf-SQUID  is made of a superconducting ring
interrupted by a Josephson  tunnel junction.  The system's
generalized coordinate is the total magnetic flux in the ring
$\Phi$, subjected to the potential \cite{Spiller92a}
\begin{equation}
{\hat U}(\Phi) = \frac{(\Phi-\Phi_{x})^2}{2L} - E_{J}
\cos\left(2\pi\frac{\Phi}{\Phi_{0}}\right) \, .
\end{equation}
Here, $L$ is the ring  inductance, $\Phi_{x}$ is an externally applied
magnetic  flux  to the  SQUID,  $E_{J}  =  I_{c}\Phi_{0}/2\pi$ is  the
maximum value of the Josephson energy, with $I_{c}$ being the critical
current of the junction, and  $\Phi_{0}=h/2e$ is the flux quantum. The
Hamiltonian of the system can be written as
\begin{equation}
{\hat H}_{0} = \frac{ Q^2}{2C} + U(\Phi)
\, ,
\end{equation}
where  $Q  =  -i\hbar  \partial/\partial\Phi$  is the  charge  on  the
junction  capacitance $C$.   The flux  $\Phi$ and  the charge  $Q$ are
canonically  conjugate operators  satisfying the  commutation relation
$[\Phi,Q] =  i \hbar$. A  typical potential of  this form is  shown in
Fig.~\ref{fig1}.

We  will first discuss  the case  of single  SQUID qubit  rotation. To
achieve  this  the SQUID  is  driven  by  two microwave  pulses.   The
microwave pulses are  considered as linearly polarized electromagnetic
fields with  their magnetic fields  perpendicular to the plane  of the
SQUID  ring. We  take the  angular  frequencies of  the two  microwave
fields $\bar{\omega}_{0}$  and $\bar{\omega}_{1}$ to  be near resonant
with  the   transitions  $|0\rangle  \leftrightarrow   |e\rangle$  and
$|1\rangle \leftrightarrow  |e\rangle$ respectively, where $|0\rangle$
and $|1\rangle$ are  the states of the SQUID  qubit and $|e\rangle$ is
an  excited state,  as is  shown in  figure \ref{fig1}.   The resonant
approximation can then be used to describe the dynamics of the system,
i.e.   we assume that  only the  states $|0\rangle$,  $|1\rangle$, and
$|e\rangle$  contribute   to  the   dynamics  of  the   system.   This
approximation  has already  been used  successfully in  several recent
articles  and the  system has  been termed  three-level $\Lambda$-type
SQUID  qubit \cite{Zhou02a, Amin02a,  Yang03a, Yang03b,  Yang03c}.  In
order to analyze our system further we use the interaction picture and
apply the  rotating wave approximation  \cite{Shore90,RWA}.  Then, the
Hamiltonian of the system can be expressed as
\begin{eqnarray}
\hat{H} = \frac{\hbar}{2}\Omega_{0}(t) e^{-i \Delta_{0} t}
|0\rangle \langle e| + \frac{\hbar}{2}\Omega_{1}(t) e^{-i
\Delta_{1} t} |1\rangle \langle e| + \mbox{H.c.} \nonumber \\
\label{eqn:HamT} \, ,
\end{eqnarray}
where $\Delta_{j} = \omega_{e}  - \omega_{j} - \bar{\omega}_{j}$, with
$j =  0, 1$ is  the microwave field  detuning from resonance  with the
$|j\rangle$$\leftrightarrow$$|e\rangle$        transition,       where
$\hbar\omega_{j}$ denotes the energy  of the $j$th stationary state of
the SQUID.

We assume  that $\Delta_{0} = \Delta_{1}$  such that the  system is at
two-photon   resonance.    Then,   in   the  rotating   wave   picture
\cite{Shore90}, obtained  by applying a unitary  transformation to the
Hamiltonian of Eq.~(\ref{eqn:HamT}), the  Hamiltonian of the system is
given by
\begin{equation}\label{ham2}
  \hat H'=\hbar\Delta|e\rangle\langle e|+\frac{\hbar}{2}\left[\Omega_0
    |0\rangle\langle e|+\Omega_1 |1\rangle\langle e|+ \mbox{H.c.}\right]\,.
\end{equation}
We require  that the light  pulses share the same  time-dependence but
their  peak  amplitudes   can  be  different,  and  there   can  be  a
phase-difference  between them.   Hence, the  Rabi frequencies  in the
Hamiltonian of Eq.~(\ref{ham2}) read
\begin{equation}\label{Rabi01}
  \Omega_0(t)=\Omega(t)\cos\varphi\,,\qquad
  \Omega_1(t)=\Omega(t)e^{i\eta}\sin\varphi\,,
\end{equation}
where $\eta$ and $\varphi$ are fixed angles. These two pulses define a
coupled state $|C\rangle$
\begin{equation}\label{Cdef}
  |C\rangle=\cos\varphi|0\rangle+e^{i\eta}\sin\varphi|1\rangle\,,
\end{equation}
and an uncoupled state
\begin{equation}\label{NCdef}
  |NC\rangle=-\sin\varphi|0\rangle+e^{i\eta}\cos\varphi|1\rangle\,,
\end{equation}
with respect to the microwave pulses \cite{Arimondo96}.  In this basis
the initial state $|\psi_i\rangle$ of the SQUID under consideration is
given by
\begin{equation}
  |\psi_i\rangle = \langle NC|\psi_i\rangle |NC\rangle +
  \langle C|\psi_i\rangle |C\rangle\,,
\end{equation}
and the Hamiltonian Eq.~(\ref{ham2}) reads
\begin{equation}\label{ham2b}
  \hat
  H'=\hbar\Delta|e\rangle\langle e|+\frac{\hbar}{2}\left[\Omega
    |C\rangle\langle e|+ \mbox{H.c.}\right]\,.
\end{equation}
This  Hamiltonian is  that of  a two-level  system. There  are several
analytically  solvable  models   for  two-level  systems  with  pulsed
excitation.   The  transfer  matrix  of  the general  solution  can  be
parameterized as
\begin{equation}\label{Umat}
  {\hat U}(t_f,t_i)=\left[\begin{array}{ccc}
      1 & 0 & 0 \\
      0 & \alpha & -\beta^{*} \\
      0 & \beta & \alpha^{*}
      \end{array}\right]\,,
\end{equation}
in the  basis $\{|NC\rangle, |C\rangle, |e\rangle \}$.  The columns of
the matrix ${\hat  U}(t_f, t_i)$ correspond to the  state vector of the
system  at time  $t_f$  if  it was  initially  in states  $|NC\rangle,
|C\rangle$, and $|e\rangle$, respectively.  Here, we need such a model
that after the pulse has passed the coupled state $|C\rangle$ acquires
a  phase shift  $-\delta$ and  the  excited state  $|e\rangle$ is  not
populated.  Hence the parameters  of the transfer matrix ${\hat U}(t_f,
t_i)$ should be given by
\begin{equation}\label{cond}
  \alpha=e^{-i\delta}\,,\qquad \beta=0\,.
\end{equation}
Therefore, at the end of the process the state of the SQUID is
given by
\begin{equation}\label{psif}
  |\psi_f\rangle = \langle NC|\psi_i\rangle |NC\rangle +
  e^{-i\delta}\langle C|\psi_i\rangle |C\rangle\,.
\end{equation}
 By  inserting  the  explicit  form  of the  scalar  products  $\langle
NC|\psi_i\rangle$ and  $\langle C|\psi_i\rangle$, and  the definitions
Eqs.~(\ref{Cdef}), (\ref{NCdef}) into Eq.~(\ref{psif}) we obtain
\begin{eqnarray}\label{psif2}
  |\psi_f\rangle \!=\!e^{-i\delta/2} \hat R_{\bm n}(\delta)|\psi_i\rangle~,
\end{eqnarray}
where    ${\bm   n}\!=\!(\sin2\varphi\cos\eta,   \sin2\varphi\sin\eta,
\cos2\varphi  )$. Apart  from a  global phase  $-\delta/2$  the states
$|\psi_i\rangle$  and  $|\psi_f\rangle$   are  connected  through  the
rotation  $\hat   R_{\hat  n}(\delta)$.   The   rotation  $\hat  R_{\hat
  n}(\delta)$  is  an element  of  the  SU(2)  group and  describes  a
rotation about the  axis $\bm n$, through the  angle $\delta$.  If the
qubit is isolated,  then the global phase $-\delta/2$  is unimportant.
If  the qubit  is part  of a  larger system,  e.g.  there  are several
qubits which form a quantum computer, then the global phase is clearly
relevant,  however, it may  be incorporated  into the  algorithm being
implemented on the quantum computer.

The simplest model for a two-level  system that can be used to
realize the dynamics described above is  the Rabi model with
rectangular pulse shape and  constant detuning. Another
possibility  is the Rosen-Zener model \cite{Rosen32,Vitanov98}
with  hyperbolic-secant pulse shape and with constant  detuning
as well.  In the case of the Rosen-Zener model  the phases are
given by rather involved formulae and we will not discuss it here.
A further, and rather general model, is the one obtained under
pulsed excitation in the case of far off-resonant Raman coupling,
i.e. the case that $\Delta \gg |\Omega(t)|/2$. Below, we describe
briefly the Rabi model and the off-resonant Raman model.  In the
Rabi model the elements of the transfer matrix (\ref{Umat}) are
given by
\begin{subequations}\label{Rabi-matrix}
\begin{eqnarray}
  \alpha &=& \left[\cos\left(\frac12 \widetilde{\Omega}T\right) +
    i\frac{\Delta}{\widetilde{\Omega}}\sin\left(\frac12
      \widetilde{\Omega}T\right)\right]e^{-i\Delta T/2},\\
  \beta  &=& -i\frac{\Omega}{\widetilde{\Omega}}
  \sin\left(\frac12 \widetilde{\Omega}T\right) e^{-i\Delta T/2},
\end{eqnarray}
\end{subequations}
where  $\widetilde{\Omega}=\sqrt{\Omega^2+\Delta^2}$.  For pulse  area
$\widetilde{\Omega}T  =  2\pi  m$,  with  $m$ being  an  integer,  the
transition   amplitude   $\beta$   is   zero,  as   is   required   in
Eq.~(\ref{cond}). Hence the rotation angle $\delta$ is given by
\begin{equation}
  \delta=\left(\frac{\Delta}{\widetilde{\Omega}} +1\right)m\pi\,.
\end{equation}
We note that the actual physical  system  is  described   by the
Hamiltonian (\ref{ham2}). The  two Rabi frequencies $\Omega_0$
and $\Omega_1$ are derived from the same  pulse according to
Eq.~(\ref{Rabi01}), which is described  by  the Rabi  frequency
$\Omega$,  detuning $\Delta$,  and duration $T$ calculated above.

In the  off-resonant Raman  model \cite{Yang03a} state  $|e\rangle$ is
eliminated adiabatically \cite{Stroud82a,Stenholm84a} and
\begin{eqnarray}
\alpha &\approx& \exp\left[\frac{i}{4 \Delta}
\int^{t_{f}}_{t_{i}}|\Omega(t)|^2 dt \right] \, ,
\\ \beta &\approx& 0 \, .
\end{eqnarray}
Therefore, the rotation angle reads
\begin{equation}
\delta = - \frac{1}{4 \Delta} \int^{t_{f}}_{t_{i}}|\Omega(t)|^2
dt \, .
\end{equation}

\begin{figure}
\begin{center}
  \centerline{\hbox{ \psfig{figure=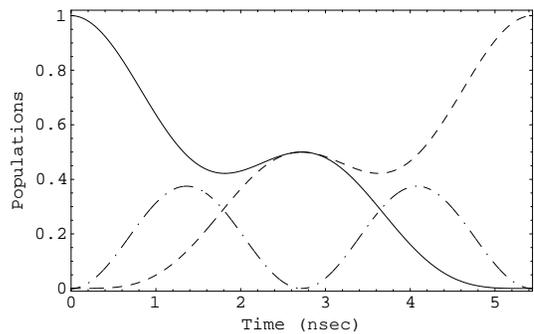,width=7cm}}}
\end{center}
\caption{ The time evolution of the populations in states
$|0\rangle$ (solid curve), $|1\rangle$ (dashed curve) and
$|e\rangle$ (dot-dashed curve) for parameters $\Omega = 2$ GHz,
$\varphi = 5 \pi/4$, $\eta = \pi$, $m = 2$, $\Delta = -2/\sqrt{3}$
GHz and $\delta = \pi$, using the Rabi model.} \label{fig2}
\end{figure}

We have performed  simulations for a realistic SQUID  system.  We used
the same parameters for the SQUID as  in the work of Zhou {\it et al.} 
\cite{Zhou02a}, i.e.  $L=100$ pH, $C=40$  fF, $I_{c} = 3.95 \mu A$ and
$\Phi_{x} =  - 0.501  \Phi_{0}$. For this  SQUID the  dissipation time
could  exceed 1 $\mu$sec  \cite{Yu02a}.  The  qubit rotation  time was
found to  be dependent on  the model of  interaction that we  used and
varied from  sub-nanosecond times  to about 30  nsec for  moderate Rabi
frequencies  values of  maximum strength  of 1-5  GHz.  An  example of
inversion from  state $|0\rangle$ to state $|1\rangle$  using the Rabi
model  is  shown in  Fig.\  \ref{fig2}.  We  note that  shorter  qubit
rotation  times can  be  achieved by  increasing  the microwave  field
intensities.   Of course, this  cannot happen  arbitrarily as  after a
certain limit the rotating  wave approximation will not be appropriate
for describing  the system and the  effect of other  states, that have
been omitted here, will have to be taken into account.

\section{Quantum information transfer and creation of entangled states}

\begin{figure}
\begin{center}
\centerline{\hbox{ \psfig{figure=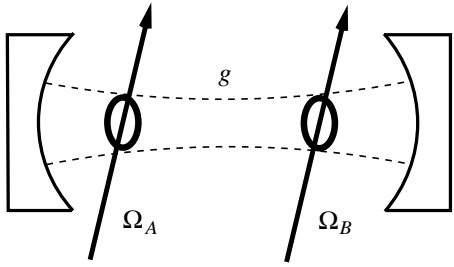,width=6cm}}}
\end{center}
\caption{A schematic representation of two rf-SQUIDs in a
  microwave  cavity.  The two  SQUIDs  are  addressed individually  by
  microwave pulses.  They are  also coupled to  the cavity  field with
  coupling strength $g$.} \label{fig3}
\end{figure}

We will  now present  a strategy for  achieving entanglement  and
also information  (coherence)  transfer  between two
$\Lambda$-type  SQUID qubits. We place both SQUIDs (we  denote
them by $A,B$) in a microwave cavity, see Fig.\ \ref{fig3},  and
assume that the transitions $|1\rangle_{j}  \rightarrow
|e'\rangle_{j}$, with $j = A,B$ and where $|e'\rangle_{j}$ are
excited states different from those used  for the single qubit
rotation, are coupled to the same cavity  mode, with coupling
constant $g$, which is assumed  real.   The   other  transitions
$|0\rangle_{j} \rightarrow |e'\rangle_{j}$,  with  $j =  A,B$
are  coupled with external  laser fields, with  Rabi frequency
$\Omega_{j}$, with $j = A,B$,  such that each microwave  field
addresses individually only one  SQUID.   We
  assume again that the cavity field and the external microwave fields
  are at two photon resonance. The Hamiltonian for the two SQUID system
  in the rotating wave picture  and in the rotating wave approximation
  is given by \cite{Yang03a}
\begin{eqnarray}\label{hamAB}
  \hat{H}_{AB}& =& \hbar\sum_{j=A,B}\Delta'|e'\rangle_{jj}\langle e'|  \\
  &{ }&+\frac{\hbar}{2}\sum_{j=A,B}\left({\Omega_j^{}}|0\rangle_{jj}\langle
  e'| + g|1\rangle_{jj}\langle e'|{\hat b}^{\dag} +
  {\rm H.c.}\right )\nonumber
\end{eqnarray}
where ${\hat b}^{\dag}$  is the creation operator of  a cavity photon.
We are  interested in  the uncoupled states  of the two  SQUID system,
which involves the  vacuum cavity state  $|0\rangle_{c}$. These
are
\begin{subequations}
\begin{eqnarray}\label{ds1}
|\psi^{I}\rangle &=& {\cal N} \bigg[ \big(\Omega^{*}_{A}(t) g
|1,0\rangle +
\Omega^{*}_{B}(t) g |0,1\rangle\big)|0\rangle_{c}  \nonumber \\
&& - \Omega^{*}_{A}(t) \Omega^{*}_{B}(t) |1,1,1\rangle \bigg] \, , \\
|\psi^{II}\rangle &=& |1,1,0\rangle \, , \label{ds2}
\end{eqnarray}
\end{subequations}
where  ${\cal  N}  = 1/\sqrt{(|\Omega_{A}|^{2}+|\Omega_{B}|^{2})g^2  +
  |\Omega_{A}|^{2}|\Omega_{B}|^{2}  }$. These states are uncoupled because
\begin{equation}
  \hat{H}_{AB}|\psi^q\rangle=0,\qquad q=I, II\,.
\end{equation}
The state of Eq.~(\ref{ds2}) is constant, it doesn't change with time.
The other  uncoupled state Eq.~(\ref{ds1})  may vary with time  as the
Rabi   frequencies  $\Omega_A(t)$  and   $\Omega_B(t)$  can   be  time
dependent.

The subset
\begin{equation}\label{H0}
  {\cal H}_0=\{|0,1,0\rangle,  |e',1,0\rangle,  |1,1,1\rangle,
  |1,e',0\rangle, |1,0,0\rangle\}
\end{equation}
of the  basis states  of the system  is closed  in the sense  that the
matrix elements of the Hamiltonian Eq.~(\ref{hamAB}) between any other
basis  state  and a  state  taken from  this  subset  is zero.   Here,
$|a,b,n\rangle  =  |a\rangle_{A}|b\rangle_{B}  |n  \rangle_{c}$  where
$|n\rangle_{c}$ denotes the state of the photons in the cavity.  Since
the uncoupled  state Eq.~(\ref{ds1}) is  a superposition of  the basis
states  from the  subset  Eq.~(\ref{H0}), then  the  dynamics of  this
uncoupled state  can be  obtain in the  subset ${\cal H}_0$.   In the
basis states of ${\cal H}_0$ the Hamiltonian (\ref{hamAB}) is given by
\begin{eqnarray}\label{hamABp}
  \hat{H}'_{AB}= \frac{\hbar}{2}\left[\begin{array}{ccccc}
      0 & \Omega_A & 0 & 0 & 0 \\
      \Omega^{*}_A & 2 \Delta' & g & 0 & 0 \\
      0 & g & 0 & g & 0 \\
      0 & 0 & g & 2 \Delta' & \Omega^{*}_B \\
      0 & 0 & 0 & \Omega_B & 0
    \end{array}\right].
\end{eqnarray}

Let us consider first information  transfer between SQUIDs $A$ and $B$
\cite{Pellizzari95a}.  If the system  is prepared initially in a state
of the form
\begin{eqnarray}\label{pts1}
|\psi(t_i)\rangle = \left(c_{0}|0\rangle_{A} +
c_{1}|1\rangle_{A}\right)|1\rangle_{B}|0\rangle_{c} \, ,
\end{eqnarray}
with  $c_{0}$  and  $c_{1}$  being arbitrary  coefficients  satisfying
$|c_{0}|^2 +  |c_{1}|^2 = 1$,  then it is  said that the state  of the
qubit $A$  is transferred in qubit  $B$ if after  some interaction the
state of the system is given by
\begin{eqnarray}\label{pts2}
|\psi(t_{f})\rangle = |1\rangle_{A}\left(c_{0}|0\rangle_{B} +
c_{1}|1\rangle_{B}\right)|0\rangle_{c} \, .
\end{eqnarray}
How can we  realize this process in the two  SQUID system?  We observe
that  the  uncoupled  state  Eq.~(\ref{ds1})  is equal  to  the  state
$|0,1,0\rangle$, for $\Omega_{B} \gg  \Omega_{A}$.  Hence let us chose
the external pulses $A$ and $B$  such that at early times around $t_i$
the  Rabi  frequencies  satisfy  the  condition  $\Omega_{B}(t_i)  \gg
\Omega_{A}(t_i)$.    Then   the    initial   state   of   the   system
Eq.~(\ref{pts1}) can be expressed as
\begin{equation}
  |\psi(t_i)\rangle = c_{0}|\psi^{I}\rangle +
    c_{1}|\psi^{II}\rangle\,.
\end{equation}
In the opposite limit  when $\Omega_{B} \ll \Omega_{A}$, the uncoupled
state  $|\psi^{I}\rangle$ coincides  with the  state  $|1,0,0\rangle$. 
Let us choose the time dependence of the pulses $A$ and $B$ such, that
at  early times  we have  $\Omega_{B}(t_i) \gg  \Omega_{A}(t_i)$, then
after a smooth  variation, at late times we  have $\Omega_{B}(t_f) \ll
\Omega_{A}(t_f)$. We also require  that the two pulses overlap.  These
conditions  ensure  that  the  uncoupled state  $|\psi^{I}\rangle$  is
well-defined throughout the whole time  evolution and the state of the
system follows adiabatically this uncoupled state \cite{Messiah}.  The
pulse sequence  described here resembles that of  the stimulated Raman
adiabatic passage (STIRAP), for reviews see \cite{Bergmann98a}. Hence,
in  the above  adiabatic  process the  state $c_{0}|\psi^{I}\rangle  +
c_{1}|\psi^{II}\rangle$ will evolve smoothly  to the final state given
by  Eq.~(\ref{pts2}), and  the required  information  transfer between
SQUIDs $A$  and $B$  will be realized  in this way.   The adiabaticity
conditions are given by
\begin{equation}\label{adicond}
  |\langle{\psi}_k|\dot{\psi}^{I}\rangle|\ll |\varepsilon_k|\,,\qquad k=1\ldots
  4\,,
\end{equation}
where   $|\psi_k\rangle$  is  an   instantaneous  eigenstate
of  the Hamiltonian (\ref{hamABp}),  and $\varepsilon_k$ is  the
corresponding non-zero eigenvalue
\begin{equation}
  {\hat H}'_{AB}|\psi_k\rangle=\varepsilon_k|\psi_k\rangle\,.
\end{equation}
The fifth eigenstate is $|\psi^{I}\rangle$ belonging to the
eigenvalue zero. The eigenvalues $\varepsilon_k$ are given by
\begin{equation}\label{eigval}
  \varepsilon_k = \frac{\hbar\Delta}{2}\pm \frac{\hbar}{2} \sqrt{
    4\Delta^2 + 2{\tilde\Omega}^2\pm 2\sqrt{(|\Omega_A|^2-|\Omega_B|^2)^2 +
  4g^4}}\,,
\end{equation}
with    ${\tilde\Omega}^2=2 g^2+|\Omega_A|^2+|\Omega_B|^2$,
and   the eigenvectors read
\begin{equation}\label{eigvec}
  |\psi_k\rangle = \frac{1}{{\cal N}_k}\left[\begin{array}{c}
      g^2 \Omega_A \\
      2 g^2 \varepsilon_k \\
      -g ( |\Omega_A|^2+4\varepsilon_k(\Delta-\varepsilon_k)) \\
      -2\varepsilon_k (g^2+ |\Omega_A|^2+4\varepsilon_k(\Delta-\varepsilon_k))\\      -\Omega_B(g^2+ |\Omega_A|^2+4\varepsilon_k(\Delta-\varepsilon_k))
      \end{array}\right]\,.
\end{equation}
By inserting Eqs.~(\ref{ds1}), (\ref{eigval}), and (\ref{eigvec})
into Eq.~(\ref{adicond})  the fulfillment of  adiabaticity can
be verified for the  chosen external  Rabi frequencies, cavity
coupling strength, and  detuning: similarly  to  the standard
three-level STIRAP,  large pulse   areas  ($\Omega_{A,B}T,
gT\gg   1$,  with   $T$  being   the characteristic length of the
pulses), and smooth, slowly varying pulse envelopes are required.

We now address  the question of creating entanglement  between
the two SQUID systems. The entangled state is defined by
\begin{equation}\label{ents}
|\psi(\theta, \xi)\rangle = (\cos\theta|1,0\rangle +
e^{-i\xi}\sin\theta|0,1\rangle)|0\rangle_c \, ,
\end{equation}
where $\theta$  and $\xi$ are fixed  angles.  In order  to obtain this
state,    we     apply    the    method     of    fractional    STIRAP
\cite{Marte91,Weitz94a,Weitz94b,Vitanov99a}.   We  chose  the  initial
state of the system to be $|\psi(t_i) \rangle = |0,1,0\rangle$ and the
order of  the pulses  as before, i.e.   at early  times ($t\rightarrow
t_i$) $\Omega_{B}(t)  \gg \Omega_{A}(t)$, so that  the uncoupled state
$|\psi^{I}\rangle$ coincides with the initial state of the system.  If
the evolution  is adiabatic and the  two pulses are  switched off such
that  $\Omega_{A}(t)  =   \Omega(t)\cos\theta$  and  $\Omega_{B}(t)  =
\Omega(t)\sin\theta  e^{i\xi}$  as  $t  \rightarrow t_{f}$,  then  the
entangled state Eq.~(\ref{ents}) is created. Moreover, by means of the
inversion of the state of SQUID $A$, the entangled state
\begin{equation}\label{ents2}
|\phi(\theta, \xi)\rangle = (\cos\theta|0,0\rangle +
e^{-i\xi}\sin\theta|1,1\rangle)|0\rangle_c \, ,
\end{equation}
can also be generated.

Our information transfer and  entanglement scheme possesses the
merits of both SQUID cavity  QED quantum computing schemes
\cite{Yang03b} and adiabatic  evolution transfer  schemes
\cite{Bergmann98a}. The  major advantages  of SQUID  cavity QED
schemes  are the  simplicity of  the coupling between the two
SQUIDs via the cavity mode, the protection of SQUIDs  from  the
interaction  with  the  environment  and  thus  the reduction  of
decoherence,  and  the technical  simplicity of  placing SQUID in
cavities. The  main advantage of adiabatic evolution transfer
methods, such as STIRAP, is  the robustness of the method for
moderate fluctuations of the microwave pulse parameters. In
addition,  as long as  the evolution remains adiabatic  the
excited states  $|e'\rangle_{j}$, with  $j =  A, B$  are
minimally populated, while the cavity  mode is only populated in
the transient regime. The latter can  be greatly suppressed,  or
completely avoided,  by keeping the coupling $g$ greater  than
the Rabi frequencies $\Omega_{j}$, with $j = A,B$.  Therefore
dissipation and decoherence can be reduced using this scheme.

\begin{figure}
\begin{center}
\centerline{\hbox{ \psfig{figure=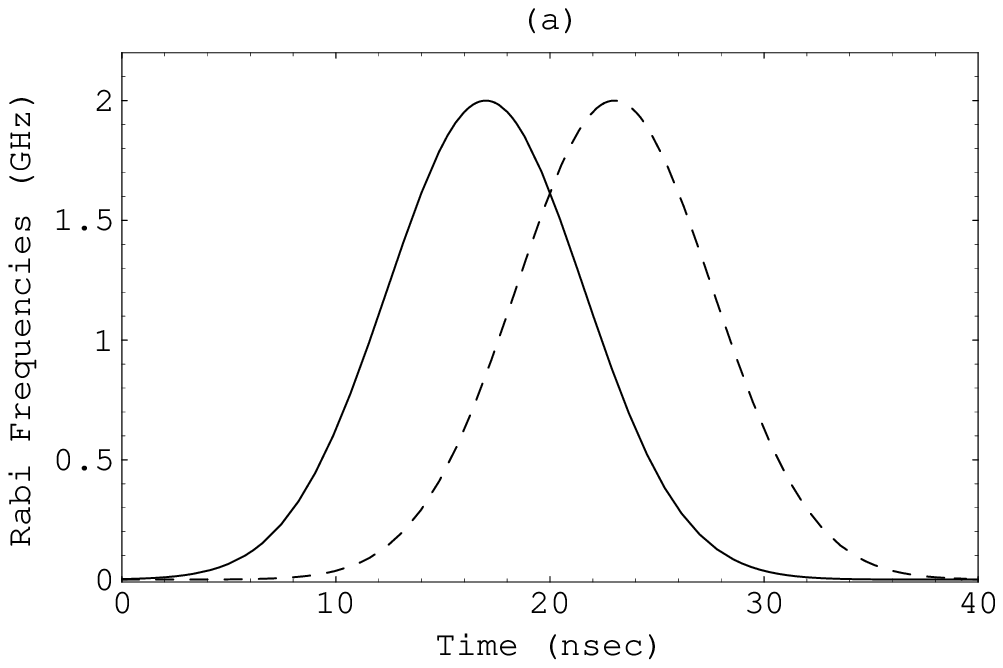,width=7cm}}}
\centerline{\hbox{ \psfig{figure=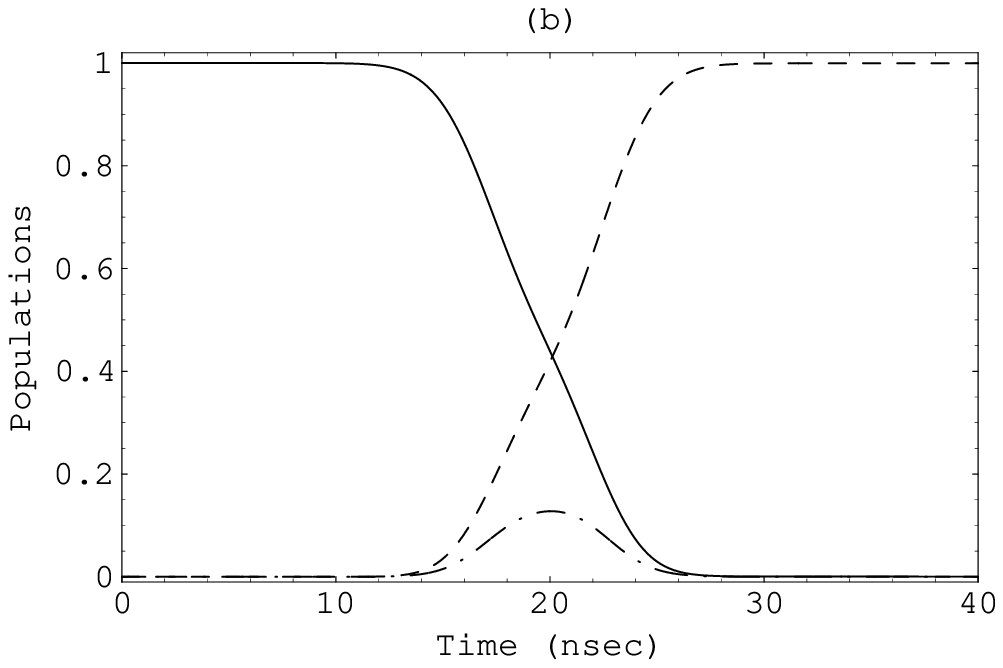,width=7cm}}}
\end{center}
\caption{(a) The Rabi frequencies $|\Omega_{A}(t)|$ (dashed curve)
and $|\Omega_{B}(t)|$ (solid curve)  for the case that
$\Omega_{A}(t) = \bar{\Omega} e^{-(t-\tau_{A})^2/\tau^{2}_{p}}$,
$\Omega_{B}(t) = \bar{\Omega} e^{-(t-\tau_{B})^2/\tau^{2}_{p}}$ for
parameters $\bar{\Omega} = -2$ GHz, $g = 3 $ GHz,
$\tau_{A} = 23$ ns, $\tau_{B} = 17$ ns, $\tau_{p} = 6.5$ ns. (b)
The time evolution of the populations in states $|0,1,0\rangle$
(solid curve), $|1,0,0\rangle$ (dashed curve) and $|1,1,1\rangle$
(dot-dashed curve) for the above parameters.} \label{fig4}
\end{figure}

\begin{figure}
\begin{center}
\centerline{\hbox{ \psfig{figure=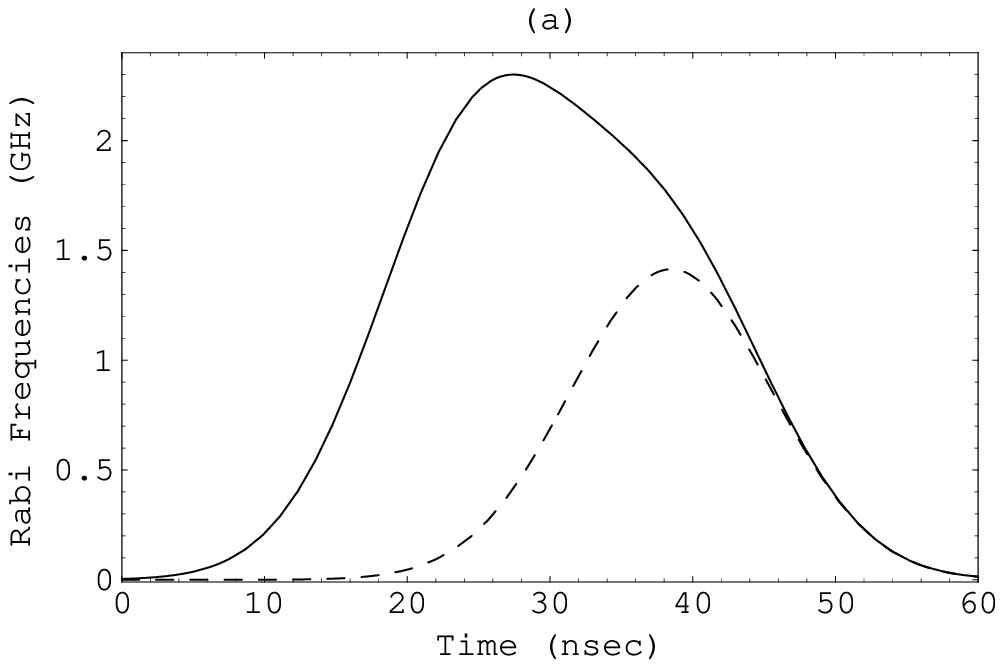,width=7cm}}}
\centerline{\hbox{ \psfig{figure=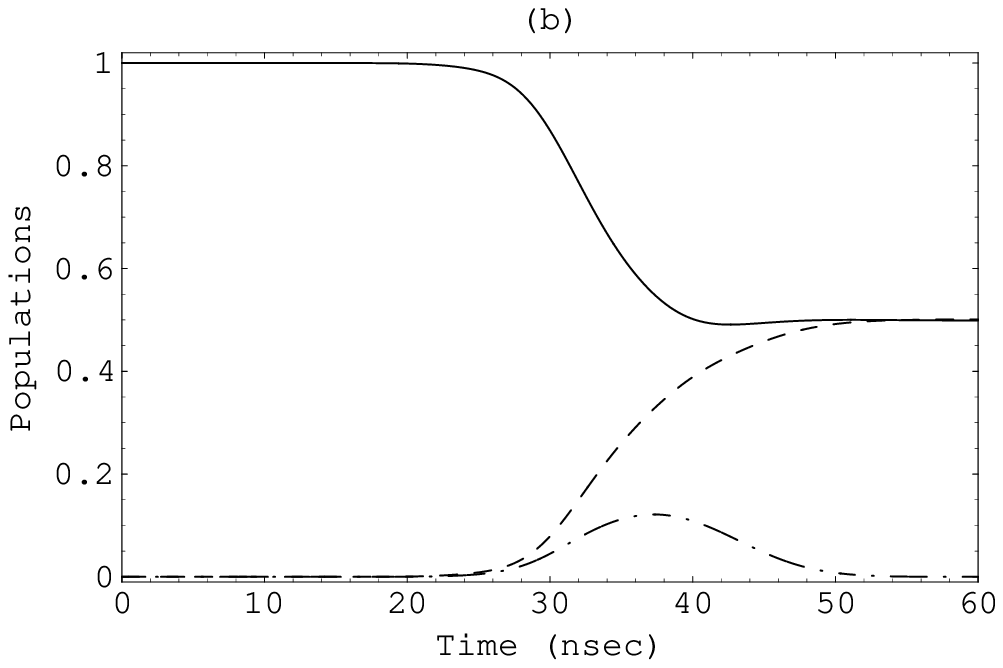,width=7cm}}}
\end{center}
\caption{ (a) The Rabi frequencies $|\Omega_{A}(t)|$ (dashed
  curve)  and  $|\Omega_{B}(t)|$  (solid  curve)  for  the  case  that
  $\Omega_{A}(t)           =          \bar{\Omega}          \sin\theta
  e^{-(t-\tau_{A})^2/\tau^{2}_{p}}$,   $\Omega_{B}(t)  =  \bar{\Omega}
  \left[e^{-(t-\tau_{B})^2/\tau^{2}_{p}}          +         \cos\theta
    e^{-(t-\tau_{A})^2/\tau^{2}_{p}}\right]$       for      parameters
  $\bar{\Omega}  = -2$  GHz,  $g =  3  $ GHz,  $\tau_{A}  = 38.5$  ns,
  $\tau_{B} = 25$ ns, $\tau_{p} = 10$ ns, $\theta=\pi/4$. (b) The time
  evolution  of  the  populations  in  states  $|0,1,0\rangle$  (solid
  curve),   $|1,0,0\rangle$   (dashed   curve)   and   $|1,1,1\rangle$
  (dot-dashed curve) for the above parameters.} \label{fig5}
\end{figure}

We have performed simulations for the SQUID presented in section
II. For coupling strengths for the Rabi frequencies and the
cavity coupling coefficient of the order of 1-5 GHz information
transfer takes from 10-30 nsec while the creation of entangled
states needs slightly larger times up to 50 nsec. A typical
example for switching from state $|0,1,0\rangle$ to state
$|1,0,0\rangle$ is shown in Fig.\ \ref{fig4}. We also present the
creation of the entangled state $1/\sqrt{2}\left(|0,1,0\rangle +
|1,0,0\rangle\right)$ in Fig.\ \ref{fig5}.

\section{Conclusion}

In summary,  we have presented schemes  for basic
state-manipulations of a  single rf-SQUID and two  rf-SQUID
qubits.  In the  first part of the  paper we have  considered a
scheme for  arbitrary rotation  of a three-level rf-SQUID qubit.
The  rotation is performed by irradiating the  SQUID  with  two
microwave  pulses that  share  the  same  time dependence. These
pulses define a coupled and an uncoupled state out of the two
SQUID states forming the qubit,  and the rotation  of the qubit
results from the phase-shift  on the coupled state caused by the
applied microwave pulses. The main advantage of our proposed
scheme is that it  can be implemented  for a large variety of
pulse  shapes and detunings, according to the capabilities of the
experimentalists.

In  the  second  part  of  the  paper we  have  proposed schemes
for information transfer and creation of entangled states between
two rf-SQUIDs. These  processes  are  mediated  by  a microwave
cavity, so that  the SQUIDs communicate via photon exchange
through a cavity mode. We have applied an  adiabatic population
transfer scheme, therefore  our method  is robust with  respect
to the moderate fluctuations of  the experimental parameters. For
proper choice of  the external Rabi  frequencies the excitation
of  the cavity field can be almost completely suppressed, hence
the decoherence due  to the imperfection  of the cavity  can be
minimized.

In closing,  we note  that our  schemes are quite  general and
can be applied to other areas of quantum computation where
three-level qubits are used such  as, for example, in atoms and
trapped ions in cavities \cite{Pellizzari95a,CQED}.

\section{Acknowledgments}

This  work  was  supported  by  the Centre  of  Excellence  Programme,
Contract  No.  ICAI-2000-70029EU.  ZK  acknowledges support  from  the
J\'anos Bolyai program of the  Hungarian Academy of Sciences, and from
the Research  Fund of the  Hungarian Academy of Sciences  (OTKA) under
contract  T43287. E.P.  acknowledges the  support of  the postdoctoral
program of the Greek State Scholarship Foundation (IKY).

\pagebreak

\vspace*{3.cm}

\begin{thebibliography}{99}
\bibitem{RWA} We note that in making the rotating wave approximation
we first keep only the driving of the transition $|0\rangle
\leftrightarrow |e\rangle$ by the microwave field with angular
frequency $\bar{\omega}_{0}$ and the transition $|1\rangle
\leftrightarrow |e\rangle$ by the microwave field with angular
frequency $\bar{\omega}_{1}$. The driving of all the other
transitions by the two fields are omitted. In addition, we omit
the usual counter-rotating terms (terms that oscillate with
frequencies $\omega_{e} - \omega_{0} + \bar{\omega}_{0}$ and
$\omega_{e} - \omega_{1} + \bar{\omega}_{1}$) but also the terms
containing the permanent matrix elements $\langle
n|\Phi|n\rangle$, with $n =0,1,e$ which exist due to the asymmetry
of the potential. The validity of this approximation for our
system has been tested with numerical simulations.
\end{thebibliography}
\end{document}